\documentstyle[aps,prb,amstex,epsfig,twocolumn]{revtex}

\begin{document}

\title{A quantum Monte Carlo study of the
one-dimensional ionic Hubbard model}

\author{Tim Wilkens and Richard M. Martin }
\address{Department of Physics and Materials Research Laboratory, University of Illinois, Urbana, Illinois 61801}
\date{\today}
\maketitle
\begin{abstract}
Quantum Monte Carlo methods are used to study a quantum phase
transition in a $1$D Hubbard model with a staggered ionic
potential ($\Delta$). Using recently formulated methods, the
electronic polarization and localization are determined directly
from the correlated ground state wavefunction and compared to
results of previous work using exact diagonalization and
Hartree-Fock.  We find that the model undergoes a thermodynamic
transition from a band insulator (BI) to a broken-symmetry bond
ordered (BO) phase as the ratio of $U/\Delta$ is increased. Since
it is known that at $\Delta = 0$ the usual Hubbard model is a
Mott insulator (MI) with no long-range order, we have searched
for a second transition to this state by (i) increasing U at fixed
$\Delta$
%TJ2
 and
%TJ2
 (ii) decreasing  $\Delta$ at fixed U.  We find no
transition from the BO to MI state, and we propose that the MI
state in $1D$ is unstable to bond ordering under the addition of
any finite ionic potential $\Delta$.  In real 1D systems the
symmetric MI phase is never stable and
 the transition is from a symmetric BI phase
to a dimerized BO phase, with a metallic point at the transition.
\end{abstract}

\section{Introduction}

Strongly-correlated systems of interacting electrons lead to many
of the most interesting phenomena observed in solid state
physics\cite{Imada}. As a function of the interaction strength,
there can be quantum phase transitions \cite{Imada} characterized
by an order parameter with the possible development of long-range
order and a transition to a broken symmetry state. Interactions
can also lead to ``Mott insulators'' (MI) and to metal-insulator
transitions \cite{MetInsTran}.  An important question is whether
or not in the thermodynamic limit a Mott insulator must be
associated with a phase transition that is accompanied by a
broken symmetry and a corresponding order parameter. In his
original work, Mott \cite{Mott49} argued that the insulating
character did not depend upon an order parameter.  On the other
hand, Slater \cite{Slater} emphasized the relation of the
insulating behavior to the long range order, and in many cases it
is known that the MI state must be accompanied by a broken
symmetry\cite{Shankar}.

To address such issues theoretically we must have methods that can
clearly distinguish metals from insulators, i.e., the ability to
transport charge\cite{griffiths,landau,kohn} vs. localization of
the electrons\cite{kohn}. Insulators at absolute zero can not
transport arbitrary amounts of charge macroscopic distances across
their bulk; however, the center of electronic charge can shift in
response to external fields, which is described in terms of
changes in polarization\cite{griffiths,landau}. The polarizability
is characterized by the degree of electronic
delocalization\cite{kohn} which increases with the proximity to
the metallic state. Recently, there have been new developments
defining macroscopic polarization and localization in terms of
the insulating ground state
wavefunction\cite{Kingsmith,OrtizMartin,RestaP,RestaL,RestaSorella,Aligia,Souza}.
These theories formulate the polarization and localization in
terms of Berry's phases \cite{berry} which can be calculated using
``twisted boundary conditions'' or in terms of the expectation
value of an exponentiated operator.  Such twisted boundary
conditions have been applied in the past to study metals and
approach metal-insulator transitions from the metallic
side\cite{IntTwistBoundCond,Scalapino,BeyersYang,kohn}. With the
recently developed methods for insulators, there are now
complementary tools\cite{Souza} to provide quantitative
information on the divergence of the localization length as one
approaches the metal-insulator transition from the insulating
side.

Generalized Hubbard models \cite{Hubbard,HubModBook} are
well-suited for studies of fundamental issues regarding metals and
insulators because they are simple models that exhibit a wide
range of behaviors depending upon the parameters in the models.
The simplest of all, the original Hubbard model with an on-site
interaction $U$ and nearest neighbor hopping $t$ in $1D$, was
solved exactly by Lieb and Wu~\cite{Lieb}.  Their paper, entitled
``Absence of Mott Transition ...'', conveys the point that there
is no change of spatial symmetry and no phase transition at any
positive U. At half-filling the model is metallic  at $U=0$,
whereas at any positive interaction U a gap exists to charge
excitations but no gap exists to spin excitations.  This is
commonly referred to as the MI state, but in this case there is
no ``Mott transition''. At any other filling, the model is always
metallic. There is never a state that would be called an ordinary
band insulating (BI) state.  However, in systems of higher
dimensionality ($d\geq 2$), a MI state is always accompanied by a
broken symmetry\cite{Shankar}.

Many new possibilities emerge for generalized Hubbard models in
$1D$. The ionic $1$D Hubbard model with two inequivalent sites,
proposed by Nagaosa\cite{Nagosa} and later by Egami\cite{Egami} as
a model ferroelectric, is ideal for studying how quantized
particle transport is modified by electron correlation in a many
body system.  On general grounds we expect a transition to occur
from an ionic band insulator to a strongly correlated Mott
insulator as $U$ is increased. Evidence for such a transition was
found in exact-diagonalization
calculations\cite{RestaSorella,OrtizMartin}, where the electronic
polarization was found to jump abruptly between two discrete
values fixed by the existence of two centers of inversion at the
two sites. Such behavior has been termed a ``topological
transition'' \cite{Aligia} that occurs in finite systems and
therefore is distinct from a true quantum phase transition. These
solutions predict that the model has a metallic point separating
two insulating phases and that a ferroelectric polarization
results only if the atomic sites are displaced from the centers of
inversion.

However, recently Fabrizio, et al.,\cite{Fabrizio} have proposed
that this model will instead exhibit two quantum phase
transitions: one from a BI state to a long range bond ordered (BO)
state, predicted to be in the Ising universality class, and a
second from the BO to the MI state, predicted to be a
Kosterlitz-Thouless transition. Such transitions to BO states have
recently been found in $1$D Hubbard models with extended
interactions (U-V) by Nakamura.\cite{Nakamura}  The BO state is a
broken symmetry state in which the system becomes ferroelectric
due strictly to electron-electron interactions even if all the
atoms are at centers of inversion.

During the course of the present work, two preprints have
reported calculations of charge and spin gaps in the
model\cite{DMRG1,DMRG2}.  Even though each work uses the density
matrix renormalization group (DRMG) that allows studies of very
large 1D systems, each group reports great difficulty in
extrapolating to large size the small spin gaps and the two
papers come to opposite conclusions regarding the existence of
the BO state.

The purpose of this paper is to study the ionic Hubbard model
using a method that (i) will treat electron correlation exactly
and (ii) scale to large systems needed to treat systems near
second-order phase transitions. For these reasons we use quantum
Monte Carlo \cite{MitasRMP}(QMC) which in principle is exact
since there is no ``fermion sign problem'' in this particular
$1$D model (so long as there is non-zero overlap between our
trial function and the true ground state). To our knowledge this
is the first QMC study of polarization and localization in any
system, and the first study of the ionic Hubbard model with
systems large enough to determine quantitatively the nature of
the transitions and whether or not there exists the spontaneously
bond-ordered phase proposed by Fabrizio, et al.\cite{Fabrizio}.
Furthermore, if there are indeed quantum phase transitions in the
ionic model -- whereas it is known that there are none in the
usual non-ionic Hubbard model -- then it follows that one must
address the issue: Is a critical degree of ionicity required, or
is the usual Hubbard model unstable to infinitesimal ionic
perturbations? It is known
\cite{Ukrainski,Baeriswyl,Hayden,Horsh,Hirsh} that the usual
Hubbard model is unstable to dimerization at all $U$. Thus a
second question is: does this instability play a fundamental role
in stabilizing the bond-ordered state?

The organization of the paper is as follows.  In Sec.II, we
introduce the model studied in this paper.  In certain cases,
depending upon the parameters of the Hamiltonian, this model is
exactly soluble.  We discuss the relevance of these solutions to
the more general case studied in this paper.  In Sec. III formulas
for evaluating the electronic polarization and localization are
presented.
%These formulas are valid only for evaluation using a
%wavefunction that adopts periodic boundary conditions.
In Sec IV, we introduce the Quantum Monte Carlo (QMC) methods
employed to evaluate expectation values and we describe their
respective limitations. These are Variational and Green's
Function Monte Carlo algorithms and the
 ``forward walking'' method for computing
expectation values of operators that do not commute with the
Hamiltonian. Our results are presented in Sec. V and comparisons
are made with previous studies using exact diagonalization and
Hartree-Fock.
%In particular, we address the nature of the quantum phase
%transition and show that the transition is
%characterized by a change of phase to a bond ordered state rather than
%to a Mott insulating state.  The latter is searched for
%- but not found - using two approaches.
In Section VI, we discuss the differences between our results and
previous studies and the consequences of our new findings.

\section{The Model}
\label{model}

The generalized ionic Hubbard Hamiltonian \cite{Nagosa} is
defined by
\begin{equation}
\label{Hamiltonian} \hat{H}=\hat{H}_{o}(t_{o},U)+\hat{H}_{\rm
Ion}(\rm \Delta)+\hat{H}_{Dim}(x),
\end{equation}
where $\hat{H}_{o}$ is the Hamiltonian of the usual Hubbard model
\begin{eqnarray}
\hat{H}_{o}(t_{o}& & ,U)= \nonumber \\
& &
\sum_{i,\sigma}\;t_{o}(c_{i+1,\sigma}^{\dag}c_{i,\sigma}+c_{i,\sigma}^{\dag}c_{i+1,\sigma})
+ U\sum_{i=1}^{L}\hat{n}_{i,\sigma}\hat{n}_{i,-\sigma}.
\end{eqnarray}
Here $c_{i,\sigma}^{\dag}(c_{i,\sigma})$ creates (destroys) an
electron of spin $\sigma$ on site $s$ while
$\hat{n}_{i,\sigma}=c_{i,\sigma}^{\dag}c_{i,\sigma}$ is the
density operator of electrons of spin $\sigma$ on site $i$. This
system is an idealized model of a chain of atoms that can have at
most $2$ electrons of opposite spin per atom.  The magnitude of
the matrix element ($t_{o}$) controls the strength of covalency in
the centrosymmetric lattice and determines the width of the energy
band in the non-interacting limit.  Interactions are included only
for electrons that occupy the same site, and the strength of
electronic correlation is determined by the ratio of $U/t_{o}$.

 The ionic term,
\begin{eqnarray}
\hat{H}_{\rm Ion}(\Delta)=\Delta
\sum_{i,\sigma}(-1)^{i}\hat{n}_{i,\sigma},
\end{eqnarray}
consists of an on-site energy($\pm\Delta$) that alternates between
neighboring sites, which is intended to model the electrostatic
potential of cations and anions in an ionic material. By adjusting
the ratio of $t_{o}/\Delta$, we can vary the degree of covalency
and ionicity to levels similar to those of real insulating
systems.

Although we will not study dimerization, {\it per se}, it is
crucial to include a dimer term that breaks the inversion symmetry
and is defined with the Su-Schrieffer-Heeger
form\cite{SuSchriefferAnsatz}
\begin{equation}
\hat{H}_{\rm Dim}(x)= \delta \: \hat{B}.
\end{equation}
Here $\delta=\alpha x$ denotes a dimerization term in the
Hamiltonian ($t_i = t_{o}(1 + (-1)^i \delta)$) that incorporates
the effect of alternately displacing the atoms $\pm x$ from their
equilibrium positions ($R(i)_{o} = ia$) and $\alpha$ is the linear
electron phonon coupling constant. The operator $\hat{B}$ is the
``bond order'' operator
\begin{eqnarray}
\hat{B}&=&\frac{2}{N}\sum_{i} (-1)^{i}\; \hat{B}_{i} \nonumber \\
&=&
\frac{2}{N}\sum_{i}(-1)^{i}\;[\sum_{\sigma}(c_{i+1,\sigma}^{\dag}c_{i,\sigma}+c_{i,\sigma}^{\dag}c_{i+1,\sigma})],
\label{Dimerization}
\end{eqnarray}
which is a staggered hopping operator, the expectation value of
which is the average difference in kinetic energy associated with
the two bonds in a unit cell.  Here $N$ is the number of sites,
$N/2$, the number of cells, and $\hat{B}_{i}$ the strength of the
$i^{th}$ bond. (Fabrizio, et al., refer to this as a
``dimerization'' operator; however, we will use the term ``bond
order'' \cite{Nakamura}, since it denotes a property of the
electronic state and may occur even if the lattice is not
dimerized.)

Exact analytic solutions for Eq.~\ref{Hamiltonian} exist in
several limiting cases. In the non-interacting case ($U=0$), the
electrons fill the lowest energy band (E(k))
\begin{equation}
E(k)=\pm[\Delta^{2}+4t_{o}^{2}\cos^{2}(k)+4(\alpha
x)^{2}\sin^{2}(k)]^{1/2}
\end{equation}
up to the Fermi k-vector $\pm\frac{\pi}{a}$.  In the case
$\Delta=x=0$ there is no gap at the Fermi surface and the system
is metallic, but for any finite $\Delta$ or $x$ a gap is opened at
the Fermi surface and the system is a band insulator. If
$\Delta=0$ and we perturb the system by adjusting $x\neq0$ the
lattice is known to suffer the famous Peierls instability
\cite{Kittel,Coulson} and energetically favors dimerization.

Exact solutions in the presence of correlation ($U\neq0$) are
restricted to cases in which (i) there is no intrasite coupling
($t=0$); (ii) there is a large displacement such that $\delta = 1$
and the lattice is completely deformed into an array of
independent dimers; or (iii) the case of the usual Hubbard model
where there is no ionic potential or lattice deformation
($\Delta=\delta=0$) for which there are exact analytic solutions
for all $U$ \cite{Lieb}. In the last case, the exact solution
predicts that at half-filling the system becomes a Mott insulator
for any non-zero $U$.  There is no change of symmetry from the
case of $U=0$ (which is a metal) and at ``very large" $U/t_{o}$
the system reduces to the Heisenberg spin model, which also has no
long range order or spin gap in one dimension.  For large $U$ the
exchange coupling of the mapped spin model is
$J=4t^{2}U/(U^{2}-4\Delta^{2})$. The MI and BI regimes are
commonly distinguished from one another in literature on the basis
of spin-charge separation \cite{Fradkin}.  In both cases there is
a gap to charge excitations but in the MI state the spin gap is
zero while in the BI state both spin and charge gaps are non-zero.

The limiting cases (i) and (ii) are also instructive for our
purposes.  In the former ($t_0 =0$) there is a transition at
$\Delta = U$ from a singlet state with two electrons on the site
with on site energy $-\Delta$, which is like a band insulator, to
a state with one electron per site which has a spin on each site
and is like a Mott insulator.  Thus one might expect a transition
from the BI state to some other phase as $U$ is increased even if
$t_0 \neq 0$.  The second case (ii) with $\delta = 1$ and
$t_{o}\neq 0$ always leads to a singlet ground state for the
isolated dimers, which relates to the known result that one has a
singlet state with a gap for both spin and charge excitations for
any degree of dimerization. Thus one can ask: does a transition
occur from the BI to MI regime as $\delta \rightarrow 0$ for
$U\neq 0$?  Is there a spontaneous\cite{Fabrizio} bond-ordered
phase? We shall test these ideas with our QMC simulations applied
to the general case where there are no exact analytic solutions.

\section{Electronic Polarization and Localization}

The issues associated with calculating the electric polarization
in an extended system have a long, torturous
history\cite{Martin1,RestaRMP}. Only recently have formulas been
devised that express the polarization and localization of
electrons directly in terms of the ground state
wavefunction\cite{Kingsmith,OrtizMartin,RestaRMP}. One type of
formulation measures the change in polarization as a Berry's phase
obtained by integrating over twisted boundary conditions and an
adiabatic parameter that characterizes the evolution of the system
as it moves from one state to another\cite{Kingsmith,OrtizMartin}.
This approach has also been extended to localization in an
independent particle formulation\cite{Marzari} and recently in a
many-body formalism\cite{Souza}.  An alternative approach has been
developed by Resta and Sorella\cite{RestaP,RestaL} and
others\cite{Aligia,Souza}, who expressed the electronic
polarization and localization in terms of the expectation value of
a complex operator
\begin{equation}
\label{ComplexRay} <\hat{Z}>=<e^{i\frac{2\pi}{L}
\sum_{i}\vec{r}_{i}}> =<\prod_{j} e^{i\frac{2
\pi}{L}\vec{r}_{j}}>,
\end{equation}
where the average is taken with respect to a truly correlated many
body wavefunction utilizing periodic boundary conditions (PBC)
sampled
%TJ2
 using one of the quantum Monte Carlo techniques discussed
in section IV.
%TJ2
 In terms of $<\hat{Z}>$ the
polarization of the many body ground state can be expressed as
\begin{equation}
<\triangle P_{el}>= \lim_{L\rightarrow\infty}\:\frac{e}{2\pi}
\:\rm Im\: \ln\:<\hat{Z}>, \label{Polarization}
\end{equation}
and  a measure of the electronic delocalization is given by
\begin{equation}
<\triangle\hat{X}^{2}>=\lim_{L\rightarrow\infty}
-(\frac{L}{2\pi})^{2}\ln|<\hat{Z}>|^{2}.\label{Localization}
\end{equation}
These expressions are  exact only in the limit of an infinitely
large system, and in practice one measures each for increasingly
larger supercells until convergence is met.  Recently Souza et al
\cite{Souza} have shown that Eqs~\ref{Polarization} and
~\ref{Localization} are in fact valid in a correlated many-body
system and related this formulation to that using twisted boundary
conditions.  They also demonstrated that the formulas relate
directly to measurable fluctuations of the polarization, thus
validating the two formulas as direct measures of electronic
polarization and delocalization.

To our knowledge the present work  is the first study of
polarization and localization on large systems with
fully-correlated many-body wavefunctions sampled using QMC.
Previous work has been limited to exact diagonalization studies on
small systems or mean field methods such as DFT and HF. For our
work we use quantum Monte Carlo methods techniques with
Eqs~\ref{Polarization} and ~\ref{Localization} because these are
directly in the form of expectation values of quantities using
wavefunctions that have the usual periodic boundary conditions.
This is a great advantage in QMC since we can use the same methods
developed for other problems \cite{MitasRMP}. The approach using
twisted boundary conditions would require a change in the
algorithms, in particular the adoption of a
``fixed-phase''\cite{BoltonFixedPhase,OrtizFixedPhase} rather than
a fixed node method. Such an approach would have important
advantages, the most significant that it would allow calculations
of polarization and localization to be done on smaller
supercells\cite{Souza}. There are other reasons that we prefer to
use the standard boundary conditions: we shall see that very large
cells are readily handled in QMC and furthermore the ability to
work with large systems is very important in conclusions on the
nature of the phase transitions in this study.

\section{Quantum Monte Carlo}
Quantum Monte Carlo (QMC) methods \cite{Hammond,MitasRMP} make it
possible to evaluate expectation values of operators in many-body
systems by stochastically sampling a probability distribution.  In
this paper we focus on two methods, Variational Monte Carlo (VMC)
and Greens Function Monte Carlo (GFMC), that can be used to
determine properties at temperature equal zero. The space of
integration ($\mathbf{R}$) is the set of all the electronic
coordinates $\{\vec{r}_{1},\ldots,\vec{r}_{N}\}$, which is sampled
by ``walkers'' which denote a set of configurations
$\mathbf{R}$.  A random walk is generated by starting from an
initial configuration $\mathbf{R}_{0}$, from which new
configurations are generated by successively
%TJ2
stepping to new random configurations,
%TJ2
 e.g., using a generalized Metropolis method \cite{Metropolis}.
This is done by accepting or rejecting new configurations at each
step based upon a chosen acceptance function ($P(\mathbf{R})$).
After a period of time the walk will stabilize such that the set
of configurations visited $\{\mathbf{R}\}$ will be distributed
according to $P(\mathbf{R})$. VMC measures expectation values by
uniformly averaging over the configurations visited by the
Metropolis algorithm where as in GFMC the average is weighted
according to $\mathbf{R}$.

\subsection{Variational Monte Carlo}

VMC measures expectation values of a variational trial
wavefunction ( $\Psi_{T}(\{\alpha\},\mathbf{R})$ ), where
$\{\alpha\}$ denotes a set of parameters that can be optimized.
Averages for an arbitrary operator $\hat{O}$ are obtained by
sampling
\begin{eqnarray}
\langle\hat{O}\rangle_{VMC} & = &
\frac{\int\Psi_{T}(\{\alpha\},\mathbf{R})\hat{O}\Psi_{T}(\{\alpha\},\mathbf{R})d\mathbf{R}}{\int|\Psi_{T}(\{\alpha\},\mathbf{R})|^{2}d\mathbf{R}}
\nonumber \\
& =
&\frac{\int|\Psi_{T}(\{\alpha\},\mathbf{R})|^{2}\hat{O}_{L}(\mathbf{R})d\mathbf{R}}{\int|\Psi_{T}(\{\alpha\},\mathbf{R})|^{2}d\mathbf{R}}
\label{VmcExpectationValue}
\end{eqnarray}
the local form of $\hat{O}_L$, defined as
$\hat{O}\Psi_{T}(\mathbf{R})/\Psi_{T}(\mathbf{R})$, over a set of
points ($\{\mathbf{R}\}$) distributed according to the modulus of
the wavefunction. The $\{\mathbf{R}\}$ are obtained by choosing
$|\Psi_{T}|^{2}$ as the acceptance function in a generalized
Metropolis algorithm.  VMC is easy to implement but is limited in
accuracy by the form of the adopted wavefunction. In our work
$\Psi_{T}$ has the Gutzwiller form\cite{Gutzwiller}
\begin{equation}
\label{trialwavefunction}
 \Psi_{Trial}(g,\Delta',\delta')=\underbrace{g^{\sum_{i=1}^{L}
\hat{n}_{i,\uparrow}\hat{n}_{i,\downarrow}}}_{Jastrow\:Term}\:D_{\uparrow}(\Delta',
\delta')\:D_{\downarrow}(\Delta', \delta'),
\end{equation}
which is a product of Slater determinants for each spin (thus
guaranteeing that the wavefunction is anti-symmetric) and a two
body Jastrow correlation function that reduces the amplitude of
configurations with doubly occupied sites for $0<g\leq 1$, thus
lowering the interaction energy.  The single body portion of
Eq.~\ref{Hamiltonian} is parameterized by $ \Delta'$ and
$\delta'$, which means the orbitals used to construct the Slater
determinants are obtained by diagonalizing the non-interacting
($U=0$) portion of the Hamiltonian ($\hat{H}(\Delta',\delta')$)
and adjusting ($\Delta',\delta'$) to optimal values that minimize
the energy in Eq.~\ref{VmcExpectationValue} wrt.
$\Psi_{T}(g,\Delta',\delta')$.

\subsection{Green's Function Monte Carlo (GFMC) for Discrete Systems}

GFMC starts with the optimized VMC wavefunction
$\Psi_{T}(g,\Delta',\delta')$ upon which a projection is applied
to obtain an improved ground state. To illustrate the principles
upon which this method depends, one can expand $\Psi_{T}$ in terms
of the eigenstates of $\hat{H}$. Then the imaginary time
propagator acting upon $\Psi_{T}$ has the form
\begin{equation}
e^{-\tau(\hat{H}-E_{0})}\Psi_{T}\;=\;e^{-\tau(\hat{H}-E_{0})}\sum_{n}C_{n}\Psi_{n}\;@>>{\tau\rightarrow\infty}>\;C_{0}\Psi_{0}\:
.\nonumber
\end{equation}
Note that the exact ground state, $\Psi_{0}$, can only be obtained
so long as it has non-zero overlap with $\Psi_{T}$.

The following is a summary of the method developed by Haaf et al
\cite{QMCLattice} some of which is used in the next section. For
lattices this projection scheme takes advantage of the fact the
spectrum of $\hat{H}$ is bound such that one can use a Green's
function projection with no finite-time-step error \cite{Koonin}
\begin{equation}
\label{ProjectionOperator}
 e^{-\tau(\hat{H}-E_{0})}\;=\;[1 - \Delta \tau ( \hat{H}
- E_{0})]^{N}|_{\stackrel{N\rightarrow\infty}{N\Delta\tau=\tau}}
\:.
\end{equation}
The propagator acting upon the trial wavefunction now
 becomes
\begin{equation}
 |\Psi^{N}\rangle\;=\;[1 - \Delta \tau ( \hat{H} -
E_{0})]^{N}\;|\Psi_{T}\rangle \: . \nonumber
\end{equation}
By inserting the identity operator in the real space configuration
basis ($\mathbf{R}=\{r_{1,\uparrow},\ldots,r_{n,\downarrow}\}$)
\begin{equation}
\sum_{\mathbf{R}}\;|\mathbf{R}\rangle\langle\mathbf{R}| \nonumber
\end{equation}
between successive applications of the projection operator and
multiplying both sides by $\langle\mathbf{R_{N}}|$
\begin{eqnarray}
\label{IdentityInserted} \Psi^{N}& &(\mathbf{R_{N}}) =
\nonumber \\
& &
\sum_{\mathbf{R}_{N-1},\ldots,\mathbf{R}_{0}}\;\langle\mathbf{R_{N}}|[1
- \Delta \tau (\; \hat{H}-E_{0}\;)]|\mathbf{R_{N-1}}\rangle\;
\nonumber \\
&          &
\langle\mathbf{R_{N-1}}|\ldots|\mathbf{R_{0}}\rangle\;\langle\mathbf{R_{0}}|\Psi_{T}\rangle
\end{eqnarray}
we obtain an expression for the wavefunction after $N$ steps in
imaginary time.  If the time step $\Delta\tau$ is sufficiently
small $\langle\mathbf{R}|[1 - \Delta \tau ( \hat{H} -
E_{0})]|\mathbf{R'}\rangle\:>0$ and can be interpreted as a
probability.  Using this probabilistic interpretation, the sum in
Eq.~\ref{IdentityInserted} above is evaluated using Metropolis.
Multiplying and dividing by
$\langle\mathbf{R}|\Psi_{Trial}\rangle$,
Eq.~\ref{IdentityInserted} above can be importance sampled
\cite{Hammond} as
\begin{eqnarray}
\label{NthProjectedPsi} \Psi^{N}& &(\mathbf{R}_{N}) =  \nonumber \\
& &
\sum_{\mathbf{R}_{N-1},\ldots,\mathbf{R}_{0}}\Psi_{T}^{-1}(\mathbf{R}_{N})\;\prod_{i=1}^{N}G(\mathbf{R}_{i},\mathbf{R}_{i-1})\Psi^{2}_{T}(\mathbf{R}_{0}),
\end{eqnarray}
where
\begin{equation} \label{GreensFnctMatrixElement}
G(\mathbf{R}_{i},\mathbf{R}_{i-1}) =
\frac{\Psi_{T}(\mathbf{R}_{i})}{\Psi_{T}(\mathbf{R}_{i-1})}\;\langle\mathbf{R}_{i}|[1
- \Delta \tau ( \hat{H} - E_{0})]|\mathbf{R}_{i-1}\rangle.
\end{equation}
Since the $G(\mathbf{R}_{i},\mathbf{R}_{i-1})$ are not normalized
to one, they can not be interpreted directly as a probability.
This is remedied by expressing
$G(\mathbf{R}_{i},\mathbf{R}_{i-1})$ as
\begin{equation}
\label{GreensFnctExpansion}
G(\mathbf{R}_{i},\mathbf{R}_{i-1})\;=\;m(\mathbf{R}_{i},\mathbf{R}_{i-1})\;p(\mathbf{R}_{i},\mathbf{R}_{i-1}),
\end{equation}
where $p(\mathbf{R}_{i},\mathbf{R}_{i-1})$ is identified as the
probability of moving from $\mathbf{R}_{i}$ to $\mathbf{R}_{i-1}$
and given by
\begin{equation}
p(\mathbf{R}_{i},\mathbf{R}_{i-1})=|G(\mathbf{R}_{i},\mathbf{R}_{i-1})|/|m(\mathbf{R}_{i},\mathbf{R}_{i-1})|,
\end{equation}
while the weight $m(\mathbf{R}_{i},\mathbf{R}_{i-1})$ normalizes
p such that $\sum_{R_{i}}p(\mathbf{R}_{i},\mathbf{R}_{i-1})=1$ and
is
\begin{equation}
m(\mathbf{R}_{i},\mathbf{R}_{i-1})=sign(\:G(\mathbf{R}_{i},\mathbf{R}_{i-1})\:)\sum_{\mathbf{R}'}|G(\mathbf{R}',\mathbf{R}_{i-1})|
.\nonumber
\end{equation}

The nodal structure of the ground state divides configuration
space into regions in which $\Psi_{T}(\mathbf{R})$ is positive or
negative, so that $G(\mathbf{R}_{i},\mathbf{R}_{i-1})$ changes
sign upon crossing the nodal surface in configuration space.
Crossing the nodal surface by a walker causes difficulties in
Monte Carlo sampling since the weight of a walker must be
positive definite if it is to be interpreted in a probabilistic
manner. In general, one must make some approximation to remedy
this problem, by fixing the sign of $\mathbf{G}$ in the Monte
Carlo sampling; this is referred to as the "fixed node
approximation", which has been described for lattice problems by
ten Haff, et al.\cite{Haff}

In the generalized Hubbard model considered here, QMC is exact
because: (i) the only nodes of the ground state wavefunction are
the points where two electrons of the same spin cross, (ii) the
nodes are the same as in the trial function which automatically
obeys this condition, and (iii) the Monte Carlo sampling is
restricted to a nodal region in which the sign of
$G(\mathbf{R}_{i},\mathbf{R}_{i-1})$ is fixed. The last condition
is realized in the present work because each move involves only
one electron moving one site at a time; we never reach the nodal
surface since neighboring $\mathbf{R}$ in which a site is doubly
occupied by two electrons of the same spin are not allowed. Thus
our algorithm samples one nodal region (either positive or
negative) which is sufficient, since they are identical due to
the antisymmetry of the wavefunction.

 Implementation of the above method is as follows.
 A VMC calculation is performed which supplies a number of walkers $\{R\}$ initially distributed according to
 $|\Psi_{T}|^{2}$.  Each of these are then randomly walked along a
 path in configuration space using $p(\mathbf{R},\mathbf{R'})$ as
 the Metropolis acceptance function of moving from $\mathbf{R}$ to
 $\mathbf{R'}$.  Each step is weighted by
 $m(\mathbf{R},\mathbf{R'})$ such that the $i^{th}$ walker's accumulated weight is
\begin{equation}
\label{AccumulatedWeight}
w_{i}^{N}=\prod_{i=1}^{N}m(\mathbf{R}_{i},\mathbf{R}_{i-1})\:.
\nonumber
\end{equation}
Expectation values for an arbitrary operator $\hat{O}$ after $N$
projections of the green's function are measured by averaging the
weighted local form of $\hat{O}$ of each walker
\begin{equation}
\label{MixedEstimator}
\langle\hat{O}\rangle_{GFMC}=\frac{\langle\Psi_{T}|\hat{O}|\Psi^{N}\rangle}{\langle\Psi_{T}|\Psi^{N}\rangle}
= \frac{\sum_{i} O_{L}(\mathbf{R}_{N}) w_{i}^{N}}{\sum_{i}
w_{i}^{N}}\:.
\end{equation}

Averages in GFMC equal the ground state expectation value only for
those operators which commute with $\hat{H}$  because the inner
product Eq.~\ref{MixedEstimator} is a ``Mixed Estimator'' between
$\langle\Psi_{T}|$ and $|\Psi_{0}\rangle$.  Operators that commute
with $\hat{H}$ share the same eigenstates and the operator in
Eq.~\ref{MixedEstimator} can be considered to act to right on
$\Psi_{0}$, thus returning the ground state and cancelling the
normalization of the denominator.  Conversely operators that do
not commute with $\hat{H}$ have different eigenstates and thus do
not cancel the normalization of the denominator in
Eq.~\ref{MixedEstimator}.  Consequently GFMC does not produce
exact results for these operators; such expectation values will be
addressed later.

\subsection{Test of GFMC on Ordinary Hubbard Model}

\begin{figure}
\centering\begin{minipage}[H]{3.5in}
\epsfig{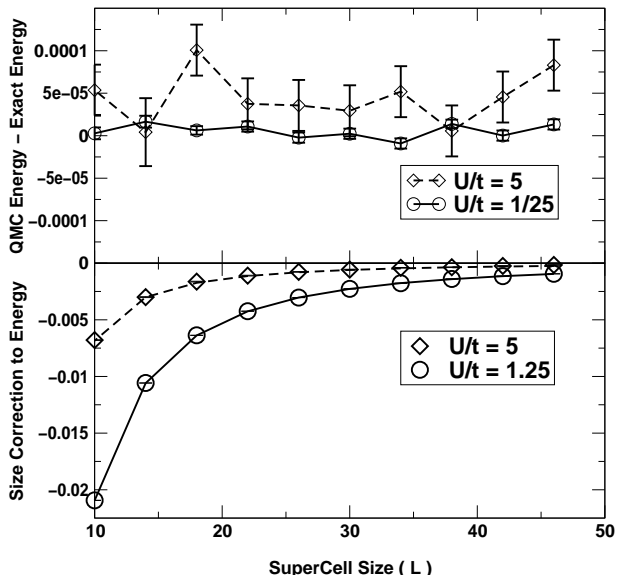}
%\end{minipage}
\caption{In the lower figure $E(4N+2)-E(\infty)$ is plotted where
$N=2,3,\ldots,11$ and infinite system estimates are those of Lieb
and Wu.  The lines are exact results from
Hashimoto\cite{Hashimoto} and the symbols are the QMC estimates
for i) $U=5$ (diamonds) and ii) $U=1.25$ (squares).  In the top
figure the energy difference between the QMC and exact results
 is plotted vs $L$.} \label{FiniteSize}
\end{minipage}
\end{figure}

The accuracy with which the energy can be measured in GFMC and the
magnitude of finite size effects can be addressed by comparing
with exact results for the usual Hubbard Model at $1/2$ filling,
which have been evaluated by Hashimoto \cite{Hashimoto} for
finite systems of $4N+2$ sites using periodic boundary
conditions. In Fig~\ref{FiniteSize} the differences in energy
between lattices of size $L$ and the thermodynamic limit is
plotted for two cases
%TJ2
$U/t=1.25 ,\; 5.0$.
%TJ2
 The finite size effects
at typical $U \approx 2.4$ are of order $0.0001 \; t$ for a
supercell of $82$ sites. Thus we do not anticipate any difficulty
in calculating the energy except in cases where there is a much
longer correlation length than in the usual Hubbard Model, e.g.,
near a phase transition where correlation lengths diverge.

\vspace{0.3in}
\vspace{0.25in}

\subsection{Stability of the Ordinary Hubbard Model ($\Delta=0$) }

For comparison to our work later it is useful also to study the
dimerized Hubbard model with $\delta \neq 0$. Work on related
issues in the past two decades has verified early theoretical
predictions \cite{Ukrainski} that electron correlation enhances
the Peierls instability of the non-interacting Hubbard model as
$\delta\rightarrow0$.  Using the Hellman-Feynman theorem the bond
order $\langle \hat{B} \rangle$ can be identified as the first
derivative of the energy wrt the lattice distortion $\alpha x$ or
$\delta$. The bond order susceptibility or the second derivative
of the energy wrt. $\delta$ has a logarithmic divergence as
$\delta\rightarrow 0$ \cite{Coulson}, which is referred to as the
Peierls instability.  The energy near $\delta=0$ varies as
\cite{Soos}
\begin{equation}
\label{Peierls} E(\delta=0) + A\delta^{\gamma}/\ln(\delta),
\end{equation}
where the amplitude $A$ and $\gamma$ are dependent upon the
strength of electron correlation.  For $U=0$ $A$ is proportional
to $t_{o}$ and $\gamma=2$, and for $U/t_{o}<<1$ variational
methods suggest the same results. In the strongly correlated
regime the lattice can be mapped onto a $1D$ Heisenberg lattice
where $A$ is proportional to $4t_{o}^{2}/U$ and $\gamma=4/3$.
Although the instability is enhanced at large U, the effect is
more difficult to observe since the electronic energy is much
smaller.

\begin{figure}
\hspace{-0.25in} \centering\begin{minipage}[H]{3.5in}
\epsfig{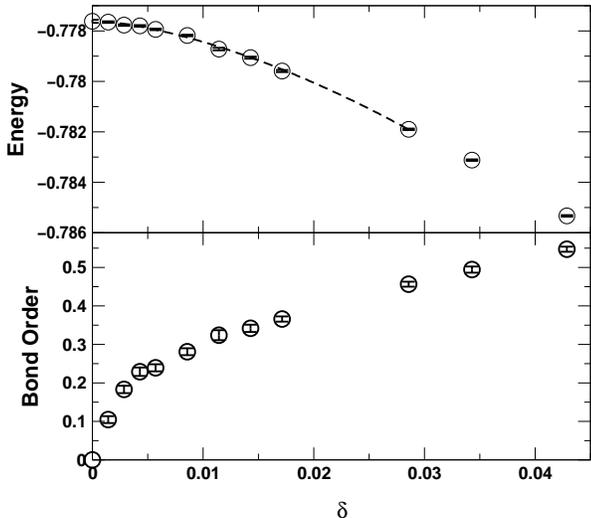}
\caption{Ground state energy and bond order vs lattice distortion
$\delta$ for $U=2.4$ in the usual Hubbard model.  The energy was
fit to the function $E(\delta=0)+A\delta^{\gamma}/\ln(\delta)$
using a non-linear least squares routine.} \label{ED24vsUPlaHub}
\end{minipage}
\end{figure}

In our studies we consider small ionic deviations ($\delta\neq0$)
from the usual Hubbard model for $U=2.4$. The QMC energy and bond
order are plotted vs $\delta$ in Fig~\ref{ED24vsUPlaHub} for an
$82$ site lattice.
%(The magnitude of finite size effects and
%the accuracy with which the energy can be measured in GFMC are
%presented in Fig~\ref{FiniteSize}, which compares with the exact
%results of Hashimoto \cite{Hashimoto} for the non-distorted
%($\delta =0$) case).
The GFMC energy was fit to Eq~\ref{Peierls} using a non-linear
least squares routine.  The parameters of the fit are
$E(\delta=0)=-0.777589(24)$, $A=1.48(17)$, and $\gamma=1.29(3)$
and give a reduced chi square of $1.58$.  This data agrees quite
will with that of Black and Emery \cite{Black} who observed
$\gamma=4/3$ in the $1D$ Heisenberg model.  The energy of the
symmetric lattice is within error bars of the exact thermodynamic
limit of $-0.77762$. The divergence of the lattice's
susceptibility of the lattice to bond ordering can be observed in
Fig~\ref{ED24vsUPlaHub}; as the level of distortion approaches
zero the bond order approaches the origin with infinite slope.

\vspace{0.2in}
\vspace{0.25in}

\subsection{Expectation Values and Forward Walking}

As noted before, GFMC does not produce exact expectation values
for operators that do not commute with $\hat{H}$.  There are
several ways to improve upon the GFMC  mixed estimator for such
expectation values. One is an approximation that is valid so long
as the VMC and GFMC averages are close to one another. Expressing
$|\Psi_{0}\rangle$ as $|\Psi_{T}\rangle+|\delta\Psi\rangle$ and
taking the inner product, the ground state expectation value can
be expressed as\cite{Hammond}
\begin{equation}
\label{GroundStateApproximation}
\langle\Psi_{0}|\hat{O}|\Psi_{0}\rangle\:\approx\:2\:\langle\hat{O}\rangle_{GFMC}-\langle\hat{O}\rangle_{VMC}+O(\delta\Psi^{2}).
\end{equation}
However, this approximation breaks down whenever the VMC trial
wavefunction is not a good approximation to $\Psi_{0}$.

The exact ground state expectation value of any operator
($\hat{O}$) can be found if the mixed expression
Eq.~\ref{MixedEstimator} is replaced by one involving the exact
wavefunction in both the bra and ket
\begin{equation}
\label{DoubleProjection} \frac{\langle\Psi_{T}|[1 - \Delta \tau (
\hat{H} - E_{0})]^{M}\hat{O}[1 - \Delta \tau ( \hat{H} -
E_{0})]^{N}|\Psi_{T}\rangle}{\langle\Psi_{T}|[1 - \Delta \tau (
\hat{H} - E_{0})]^{M}[1 - \Delta \tau ( \hat{H} -
E_{0})]^{N}|\Psi_{T}\rangle}.
\end{equation}
This can be accomplished by ``forward walking'' \cite{Hammond},
which can be simply expressed in terms of the GFMC method
previously discussed. The same methods and terminology used in
GFMC are also applicable here.  Inserting the identity operator
between each projection and using importance sampling
Eq.~\ref{DoubleProjection} can be rewritten as
\begin{eqnarray}
\sum_{\mathbf{R}_{N+M},\ldots,\mathbf{R}_{1}}\;&
&[\prod_{i=N}^{N+M-1}G(\mathbf{R}_{i+1},\mathbf{R}_{i})]\;\:O_{L}\:\;\nonumber
\\
& &
[\prod_{i=1}^{N-1}G(\mathbf{R}_{i+1},\mathbf{R}_{i})]\Psi^{2}_{T}(\mathbf{R}_{0}).
\end{eqnarray}
The $G(\mathbf{R},\mathbf{R'})$ are sampled as before in terms of
a probability function ($P(\mathbf{R},\mathbf{R'})$) and weight
($M(\mathbf{R},\mathbf{R'})$).  A series of $i$ walkers, initially
distributed according to the VMC trial function, are stepped along
paths ($\{\mathbf{R}_{i}\}$) in configuration space by Metropolis
sampling.  After $N$ projections the accumulated weight of each
$\{\mathbf{R}_{i}\}$ is the product of all steps weights, as
defined in Eq.~\ref{AccumulatedWeight}.  The walkers weights are
distributed according to the mixed probability distribution
$\Psi_{T}(\mathbf{R}_{N})\Psi_{0}(\mathbf{R}_{N})$.  The local
form of $\hat{O}$ ($O_{i}(\mathbf{R}_{N})$ is measured for each
walker but not averaged as it is in GFMC.  The walkers are moved
an additional $M$ steps in imaginary time over which they
accumulate post measurement weights ($w_{i}^{M}$). Averages are
computed using each walkers accumulated weight before and after
measuring $O_{i}(\mathbf{R}_{N})$
\begin{equation}
\label{ForwardWalkingAverages}
 \frac{\sum_{i}
w_{i}^{M}[w_{i}^{N}O_{i}(\mathbf{R}_{N})]}{\sum_{i} w_{i}^{M}
w_{i}^{N}}\:.
\end{equation}
Although this method is in principle exact, assuming the nodal
structure of $\Psi_{0}$ is known, it also has its disadvantages.
In particular, the width of the post-measurement weight
distribution grows with the number of steps $M$, thereby
increasing the fluctuation of the forward walking estimates.  To
achieve a desired level of accuracy additional measurements are
needed but the error in QMC decreases inversely with the square
root of their number. Consequently, to obtain the same error as
that in GFMC, forward walking may require many times more
estimates in Eq.~\ref{ForwardWalkingAverages}. This is the
limiting factor in applying forward walking.

\begin{figure}
\centering\begin{minipage}[H]{3.5in}
\epsfig{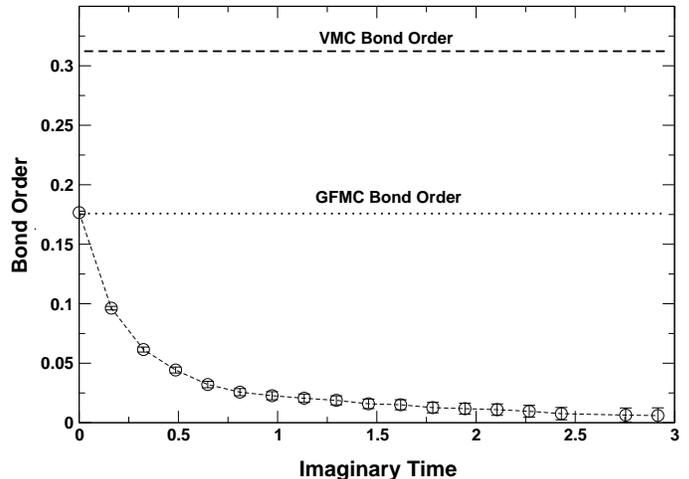}
\caption{Illustration of forward walking for $U=1.8$ and
$\{\Delta,\delta\}=\{4/7,0\}$.  The results are those obtained
using a wavefunction of the form in Eq~\ref{trialwavefunction}
with variational parameters $\{\Delta,\delta\}$ = $\{0.38,2/35\}$,
which is much worse than a typical optimized trial wavefunction
used in the present work.}
\label{ForwProjTest}
\end{minipage}
\end{figure}

To illustrate the practicality  and usefulness of this method, we
show in Fig~\ref{ForwProjTest} the bond order $<\hat{B}>$ as a
function of forward walking for the centrosymmetric lattice at
$U=1.8$, $\Delta=4/7$ and $L=62$ sites. We have chosen a poor
trial wavefunction biased towards bond ordering by defining the
determinant part of the wavefunction Eq.~\ref{trialwavefunction}
using a Hamiltonian with  $\delta ' =2/35$.  At this particular
$U$ the system is a band insulator (as shown below), consequently
the bond order must be zero; however, the average bond order in
VMC and GFMC is non-zero as a result of using this trial
wavefunction.  The VMC expectation value of the bond order is
substantially different from $0$ and reflects the poor quality of
the trail state; whereas the GFMC average is closer to the exact
result but remains far from satisfactory.  The results in
Fig~\ref{ForwProjTest} illustrate the strengths and weaknesses of
forward walking.  The key point is that there is a competition
between the improvement of the estimate and the growth of the
statistical errors with projection time. As shown in the figure,
the method vastly improves the results even for very poor
wavefunctions without the proper symmetry.  In general,  we use
much better trial wavefunctions, and so the convergence to the
exact result in our work below is more rapid than that depicted in
Fig~\ref{ForwProjTest}.

\vspace{0.3in}
\vspace{0.25in}

\section{Results for Ionic Hubbard Model}

The unit cell for the ionic Hubbard model is composed of $2$ sites
and the Hamiltonian is given in Eq.~\ref{Hamiltonian}. In order
to understand the meaning of the polarization and bond order in
this system, it is helpful to consider first the non-interacting
case with $U=0$, where one can visualize the electronic
properties in terms of Wannier functions.  At zero dimerization
($\delta=0$) the Wannier functions are centered on the sites
whose onsite energy is shifted by $+\Delta$ and $-\Delta$. The
$2$ electrons of opposite spin in each unit cell both occupy the
lowest energy Wannier function centered on the lower energy site.
In the dimerized lattice ($\delta \neq 0$), the centers are
displaced from the sites.  The magnitude of $\delta$ dictates the
amount by which they are off center from the lowest energy site
while sign of $\delta$ determines whether the center is to the
left or right of this site. Strictly speaking the existence of
Wannier functions in correlated systems ($U\neq0$) has yet to be
proven but we will continue to use the concept of the center of
the localized states for illustrative purposes. As U increases
the electrons find it energetically undesirable to occupy the
same site, and in the dimerized state the center of the
distribution shifts further away from the low energy site. The
limit of this displacement (i.e., polarization) is $P= 1/2$ since
one would never favor having more than one electron on the higher
energy site. Similarly, the limit of the bond order is $\langle
\hat{B} \rangle = 2$ corresponding to isolated dimers.

As $\delta \rightarrow 0$ there are only three possibilities. If
there is a spontaneous breaking of the inversion symmetry the
polarization can assume any fractional value between 0 and 1/2. If
there is no breaking of symmetry, there are still two
possibilities since there are two centers of symmetry: the
polarization can be 0 or 1/2.  If the reference point defined to
be zero is the usual band insulator where both electrons occupy
the Wannier function centered on the lower energy site, it has
been proposed\cite{RestaSorella,Ortiz,Gidopoulos} that P = 1/2
corresponds to a Mott insulator with no long range order.

%Model has been extensively studied in many different cases
%\Delta=0 ( long known that e-e correlation enhances charge
%transport, studied by looking at x\neq0 and allowing x\rightarrow 0.

We first report results of our study of the ionic Hubbard model
with parameters fixed at the values used in previous
work\cite{RestaSorella,Ortiz,Gidopoulos}, so that direct
comparisons can be made.  The energy scale is set by defining
$t_{o}=1$, $\Delta/t_{o} = 0.5714 $ and $\alpha a/t_{o}=40/7$. The
previous conclusions with which we will compare are based upon
exact diagonalization of the many-body Hamiltonian in small
supercells\cite{RestaSorella,Gidopoulos} and Hartree-Fock
calculations\cite{Ortiz}.  The study\cite{RestaSorella} using
exact diagonalization of $8$ site lattices with twisted boundary
conditions found a jump of $1/2$ in the electronic polarization
for $\delta = 0$, i.e. an electron in each unit cell being
transported $1/2$ lattice constant, at a critical value of $U$
($U_{c}=2.26$). This was interpreted as a transition between BI
and MI phases, which was supported by Hartree Fock (HF)
calculations that showed similar behavior at $U_{c} = 2.46$.
Extrapolations using larger cells of $12$ sites\cite{Gidopoulos}
find $U_{c}=2.86$, presumably a more converged value. The key
points are: (i) the transition point $U_{c}$ is found to be a
metallic point with divergent delocalization; (ii) effective
charges diverge and change sign at the transition; and (iii) there
is no sign of the bond-ordered state predicted  by Fabrizio, et
al.\cite{Fabrizio}. This new state would have long range order
and break the inversion symmetry of the lattice, thus allowing
the polarization to take any fractional value.

The present work is based upon the QMC algorithms described
earlier and the formulas for polarization and localization in
section III. The first step in applying the QMC methods is to find
a trial wavefunction that has as much overlap with the true ground
state as possible.  This is achieved by optimizing the parameters
$\{g,\Delta',\delta'\}$ to minimize the energy.  To determine the
optimal value of $g$ we have used a newly devised technique that
significantly reduces the amount of computational effort required
\cite{Koch}. Using the optimal Gutzwiller parameter the energy of
$\Psi_{T}(g,\Delta',\delta')$ for different $\Delta'$ and
$\delta'$ is sampled using VMC.  We adjust $\Delta'$ and $\delta'$
to lower the VMC energy and measure it at several points in the
neighborhood of its minimum.  A curve fit is then performed using
these points to determine the optimal $\Delta'$ and $\delta'$.

\subsection{Comparison with Exact Diagonalization and Hartree-Fock}

\begin{figure}
\hspace{-0.25in} \centering\begin{minipage}[H]{3.5in}
\epsfig{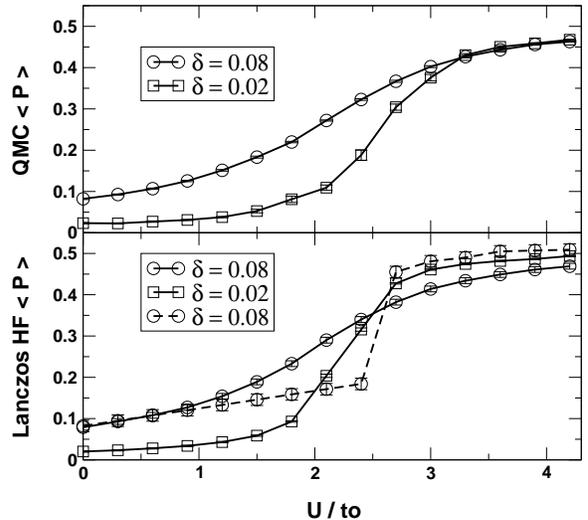} \caption{Exact
QMC measurements of polarization ( upper figure ) in comparison
with previous Lanczos and HF results ( lower figure ). Results are
illustrated for staggered transfer integrals of $1\pm0.02$ (
squares ) and $1\pm0.08$ ( circles ).  The HF results are depicted
by the dashed line in the lower figure. } \label{Fig2}
\end{minipage}
\end{figure}

Previous studies distorted the lattice by varying degrees and
these results provide a basis of comparison with QMC.  We have
measured the polarization of the ionic lattice for large
($\delta=0.08$) and small ($\delta=0.02$) lattice distortions and
plotted these with the corresponding results of previous studies
in Fig~\ref{Fig2}.  Size effects are accounted for by
extrapolating to the thermodynamic limit in $1/L$; this will be
outlined more clearly in the following section.  The Lanczos
results agree well with those of QMC for $\delta=0.08$ considering
the fact they were obtained using $8$ site supercells with twisted
boundary conditions.  This is in agreement with previous studies
using exact diagonalization \cite{Hayden,Soos} on the usual
Hubbard model which found that small cells of this size were
sufficient to reach thermodynamic convergence in the
$0.05\leq\delta\leq 0.1$ regime, whereas convergence with cell
size is worse for smaller $\delta$. (The jump in the polarization
found in HF calculations for non-zero $\delta$ is unphysical and
arises because the mean field approximation leads to an
anti-ferromagnetic ground state. In $1D$ this is strictly
prohibited because quantum fluctuations are strong enough to
destroy long range order in any continuous quantity.) As the
magnitude of the distortion ($\delta$) approaches $0$ the
difference between QMC and Lanczos becomes greater.  Studies
using exact diagonalization and HF observed that the polarization
as a function of $\delta$ tended to $0$ below a critical $U$ and
$1/2$ above it; consequently the dynamic charge was observed to
change sign upon crossing this critical point.  In the lower plot
of Fig~\ref{Fig2} this is exhibited as a crossing of the curves
for $P(\delta)$.  On the contrary we observe that the dynamic
charge remains the same sign for the entire range of $U$ studied
(except that the sign of $dP/d\delta$ is difficult to establish
for large $U$ where it is near zero). This difference is
attributed to the fact that for small $\delta=0$ and for $U$ near
$U_{c}$ the electrons are very delocalized and correlation lengths
exceed the cell size\cite{Souza}. Integrating over twisted
boundary conditions provides thermodynamically quenched
expectation values so long as the Wannier functions of the
eigenstates $\Psi_{k}$ have vanishing overlap.  Near the critical
point the $\Psi_{k}$, obtained by exact diagonalization of $8$ and
$12$ site rings, do overlap significantly; consequently,
regardless of the number of k-space points averaged over by Resta
and Sorella, the polarization will not converge to that of QMC.

\vspace{0.2in}
\vspace{0.25in}

The average polarization in Fig~\ref{Fig2} was measured using
forward walking. For the polarization at these values of $\delta$,
forward walking is not essential and the same results can be
obtained using Eq~\ref{GroundStateApproximation}. However,
localization is more sensitive to inaccuracies in the wavefunction
and accurate expectation values are provided only by using forward
projection even at these relatively large values of $\delta$. As
the level of dimerization is reduced such that $\delta\rightarrow
0$ the extrapolation technique breaks down and only forward
projection can provide accurate estimates for polarization and
localization. Consequently we only report in this paper those QMC
results obtained using forward projection.

The observation of a topological transition in the work of Resta
and Sorella, and Guidopoulos, et al., is based upon their finding
that as $\delta\rightarrow 0$ the polarization jumps
discontinuously from 0 to 1/2 at a critical $U_c = 2.26$ (or $U_c
= 2.86$).  This means that the dynamic charge ($Z$), defined as
$\partial P/\partial x|_{\delta=0}$, diverges and changes sign at
$U_c$ as $\delta\rightarrow 0$.  In latter work \cite{RestaL}
Resta and Sorella showed that $U_c$ is a metallic point where the
electronic localization length ($<\Delta^{2}X>$) diverges, and
Guidopoulos, et al., found energy gaps that extrapolated to zero.
We will compare these results with our work below.

\subsection{Phase transition to Bond-ordered State}

\begin{figure}
\centering\begin{minipage}[H]{3in}
\epsfig{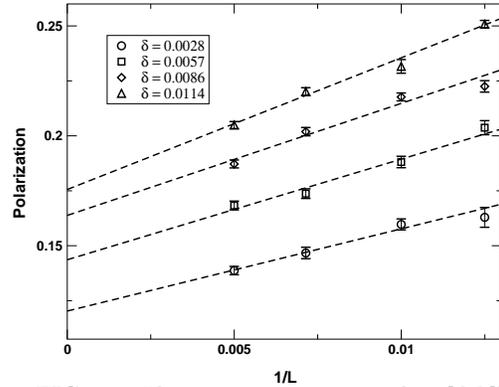} \caption{The
points represent the QMC $P$ obtained using forward walking for
system sizes of $80,\:100,\:140$ and $200$ sites for different
levels of dimerization $\delta=\{0.0028,0.0056,0.0085,0.0114\}$
in ascending order.} \label{PDiffLextrap}
\end{minipage}
\end{figure}

We have measured the forward walking estimators for $P$,
$<\hat{B}>$ and $<\Delta^{2}X>$ and taken the limit of $\delta
\rightarrow0$ to study the nature of the quantum phase transition.
The formulas used to obtain expectation values for polarization
and localization are only accurate in the limit
$L\rightarrow\infty$. This limit is taken by fitting measurements
at finite $L$ to a linear least squares fit in $(1/L)^{\gamma}$
and extrapolating to $0$.  We have found $\gamma=1$ to accurately
account for the finite size effects of $P$ and $\gamma=2$ for
$<\Delta^{2}X>$.  This scaling has only been found appropriate
upon increasing the supercell size above a critical threshold
which depends on the proximity of the metallic state.  The
accuracy of the finite size corrections to $P$ are illustrated in
Fig~\ref{PDiffLextrap} at $U=2.7$ for different magnitudes of
$\delta$.  The data in Fig~\ref{PDiffLextrap} was collected near
the critical point of the phase transition, where size effects
are large and must be treated accurately. If the system is
sufficiently far from such a critical point, size effects are
less pronounced and there is a more rapid convergence to the
thermodynamic limit.

%\vspace{-0.15in}

\begin{figure}
\hspace{-0.25in} \centering\begin{minipage}[H]{3in}
\epsfig{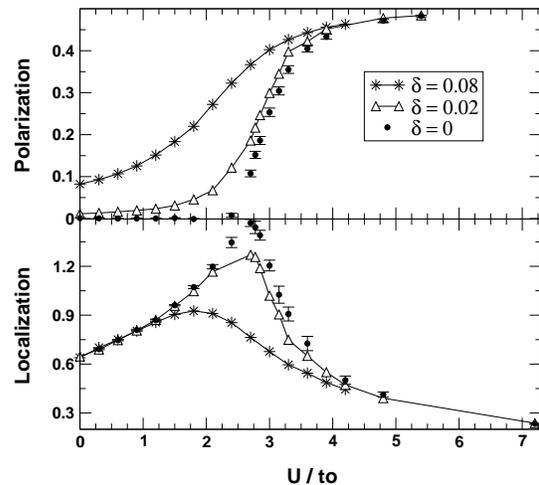}
\caption{$P(L=\infty)$ and $<\Delta^{2}X>$ for various levels of
dimerization ($\delta$).  The extrapolated centro-symmetric
polarization and localization is represented by the points with
error bars.} \label{PLvsUD5714}
\end{minipage}
\end{figure}

Using the infinite $L$ estimates for the polarization and
localization on lattices dimerized by
$\delta=\{0.0028,0.0056,0.0085,0.0114\}$ we have performed a
linear least squares fit and extrapolated to the centrosymmetric
limit ($\delta=0$).  This method makes the assumption that the
response of the lattice to dimerization is linear.  However, near
the phase transition non-linearity will cause this to break down.
In Fig~\ref{PLvsUD5714} we have plotted the polarization and
localization of the ionic model for different magnitudes of
dimerization.

%\vspace{-0.1in}
\vspace{0.25in}

\begin{figure}
\centering\begin{minipage}[H]{3in}
\epsfig{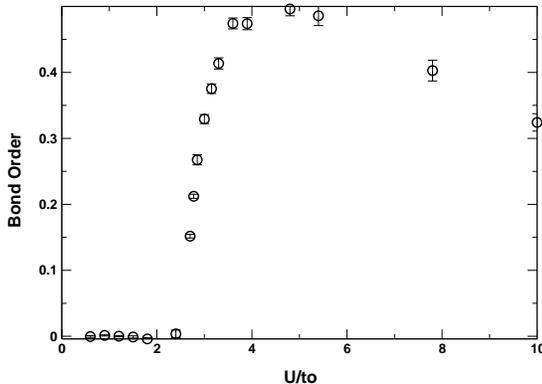} \caption{QMC Bond
Order of centro-symmetric lattice for $\Delta/t_{o}=0.5714$  and
$L=142$. Results obtained by extrapolating bond order on distorted
lattices of $\delta=\{0.0028,0.0056,0.0085,0.0114\}$ to
$\delta=0$.} \label{DimerPlot}
\end{minipage}
\end{figure}

The phase transition we observe using QMC differs from the
topological transition found using exact diagonalization and HF.
As previously mentioned, in the BI and MI phases the polarization
is restricted to $0$ or $1/2$.  Resta and Sorella identified the
shift from $0$ to $1/2$ as the signature of a $BI \rightarrow MI$
transition. However, we observe that $P$ takes a continuous range
of values in the centrosymmetric limit which cannot occur in
either the BI or the MI phase. This can only occur if the global
inversion symmetry of the lattice is broken by a long range bond
ordered state, predicted by Fabrizio et al \cite{Fabrizio} on the
basis of field theory arguments in which he mapped the Hamiltonian
onto two Ising spin models.  The order parameter of this phase
transition is the average bond order function $<\hat{B}>$, where
$\hat{B}$ is given in Eq~\ref{Dimerization}. The spontaneous bond
order of the centrosymmetric lattice is obtained by extrapolating
to $\delta=0$ the $\langle {\hat B} \rangle$ of the same distorted
lattices as before.
%for which $\delta=\{0.0028,0.0056,0.0085,0.0114\}$.
We fixed the supercell size to $142$ sites and found the
consequent size effects are within order of the error with which
we can measure the bond order.  These results are plotted in
Fig~\ref{DimerPlot}.

\vspace{0.25in}
\vspace{0.25in}

\begin{figure}
\centering\begin{minipage}[H]{3in}
\epsfig{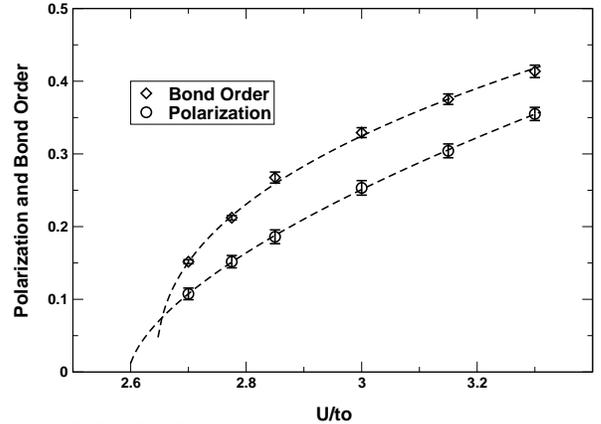} \caption{QMC
polarization and bond order near the critical point and their
relative fits to Eq~\ref{CritFit}. } \label{CriticalFit}
\end{minipage}
\end{figure}

We have attempted to classify the quantum phase transition by
fitting the polarization and bond order of the centrosymmetric
lattice to a function of the form
\begin{equation}
\label{CritFit} A[U-U_{c}]^{\xi},
\end{equation}
where $\xi$ is the critical exponent and determines the
universality class of the transition.  A non-linear least squares
routine was used to fit the data, with fitted parameters $U_{c}$,
$A$, and $\xi$ listed in Table I.  In Fig~\ref{CriticalFit} the
data for $P$ and $<\hat{B}>$ and the corresponding fits are
plotted.  Both quantities behave similarly near the critical point
and the $U_{c}$ of each is nearly identical. We find $\xi$ for
$P$ and $<\hat{B}>$ are near $1/2$, the expected mean field
exponent.   On the other hand,
 Fabrizio, et al.\cite{Fabrizio},  predicted that the transition is of the Ising
universality class and thus $\xi$ should be $1/8$. We do not know
whether the difference is real or it is simply due to the
possibility that the range of $U-U_{c}$ over which the scaling
belongs to the universality class is too small for us to observe
in the present work.

\hspace{0.25in}
\vspace{0.25in}

\begin{figure}
\centering\begin{minipage}[H]{3in}
\epsfig{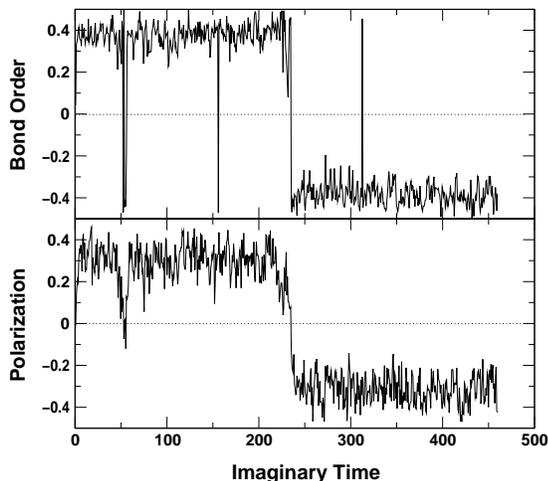}
%\end{minipage}
\caption{Illustration of the phase separation in the symmetric
case ($\delta=0$,$\Delta=0.5714$,$U=3.45$) of the polarization and bond order.
The lines
depict the measurements over which averages are obtained in QMC.}
\label{PhaseSep}
\end{minipage}
\end{figure}

Alternatively, the existence of the bond ordered state can be
observed by directly studying the symmetric lattice without any
lattice distortion. Quantum Phase transitions (QPT) are
characterized by a symmetry breaking that occurs in the
thermodynamic limit.  Below $U_{c}$ the lattice is a band
insulator with no bond order but above $U_{c}$ the electrons will
spontaneously choose to bond order with $\pm |\langle
\hat{B}\rangle|$. There are two such states characterized by the
same magnitude but opposite sign of the bond order.  For any
finite system the ground state remains a linear combination of
both.  However in the limit $L\rightarrow\infty$ one of these is
arbitrarily chosen as the ground state.  Even though the QMC
simulations of the symmetric lattice for $U>U_{c}$ measure zero
bond order and polarization for long simulations, the imaginary
time evolution of the simulations clearly depict the projected
ground state moving from one of these bond ordered states to the
other.  This phase separation gives rise to large auto
correlation times.  The evolution of the bond order and
polarization in imaginary time are illustrated in
Fig~\ref{PhaseSep} for $U=3.45$ and $L=60$ sites. As the ground
state moves between BO states of opposite symmetry both the
polarization and dimerization are observed to change sign.  This
provides an alternative method of detecting the existence of the
BO state.

\hspace{0.5in}
\vspace{0.25in}

At large $U$ one might expect that
%TJ2
 the ionic
%TJ2
$1D$ Hubbard model is a Mott Insulator.  Fabrizio, et
al.\cite{Fabrizio}, predict the existence of a
Kosterlitz-Thouless transition for large $U/\Delta$ at which the
lattice becomes a Mott Insulator.  Such a transition would be
characterized by polarization of $1/2$ and no bond order. On the
contrary we do not observe either for any $U$ considered which
included values up to $U = 10$.  The bond order does diminish but
appears to asymptotically approach $0$ and the polarization
appears to converge to $1/2$ only in the limit of
$U\rightarrow\infty$.  Yet working with such strongly correlated
systems has the disadvantages that (i) fluctuations of the local
estimators increase due to greater inaccuracies in the trial
wavefunction and (ii) forward walking works so long as the trial
wavefunction has some overlap with the exact ground state and as U
increases overlap with the exact ground state diminishes.

\subsection{Phase transition as a function of ionicity $\Delta$}

\begin{figure}
\centering\begin{minipage}[H]{3.5in}
\epsfig{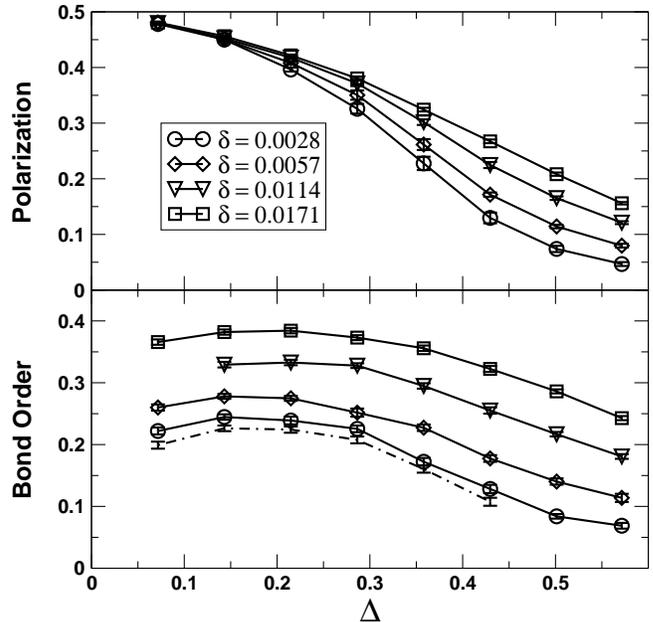} \vspace{0.2in}
%\end{minipage}
\caption{Polarization and Bond Order vs Ionic potential $\Delta$
for $\delta=\{0.0028,0.0057,0.0114,0.0171\}$ and $U=2.4$.  The
extrapolated bond order of the centro-symmetric lattice is
denoted by the dotted line.} \label{PBvsDelta}
\end{minipage}
\end{figure}

An alternative approach to study the phase transition(s) is to
diminish the ionic potential $\Delta$ while keeping $U$ fixed, so
that the ratio $U/\Delta$ increases.  We have fixed the strength
of electron correlation to $U=2.4$ and studied the bond order and
polarization for $0<\Delta\leq 0.5714$.  The behavior of the
centrosymmetric lattice is inferred using two approaches (i)
extrapolating results obtained on lattices with $\delta\neq 0$ and
(ii) looking for evidence of phase separation in the symmetric
case. Fig~\ref{PBvsDelta} shows the results of the first approach
for a fixed supercell length of $142$ sites. In the first we've
neglected size effects and fixed the super cell length to $142$
sites.  At large $\Delta$ the single body contribution to the
Hamiltonian is the dominant term and the lattice is a band
insulator. Consequently the bond order and polarization are $0$.
However, as $\Delta\rightarrow 0$ a transition occurs to a BO
state where the bond order is non-zero and the polarization
assumes values between $0$ and $1/2$ as before. (The transition is
rounded at this fixed cell length.) The bond order in the $\delta
\rightarrow 0$ limit is shown by the dotted line in
Fig~\ref{PBvsDelta}.  These results were obtained by linearly
extrapolating the bond order at finite $\delta$.  (No
extrapolation was performed for the polarization since it is
sensitive to size effects that were addressed in the previous
section.)

Our results indicate the bond-order state exists at all values of
$\Delta \neq 0$ studied.  The finite value of the bond order for
 $\delta \rightarrow 0$ shown in Fig~\ref{PBvsDelta}
contrasts sharply with the vanishing of the bond order as $\delta
\rightarrow 0$ for the non-ionic Hubbard model ($\Delta=0$) as
shown in Fig~\ref{ED24vsUPlaHub}. At $\Delta=\delta=0$, we always
find $<\hat{B}> = 0$ and polarization equal 1/2 as they must be
for a MI state with no long range order.  However, our QMC
simulations of the symmetric case ($\delta=0$ and $U=2.4$) at the
smallest value of the ionic potential studied $\Delta=0.0716$
reveal two BO states phase separating in imaginary time
qualitatively the same as shown in Fig~\ref{PhaseSep}.  Thus from
our studies, there is no sign of a second transition to a MI state
as proposed by Fabrizio, et al.\cite{Fabrizio}, and the long range
bond ordered state in Fig~\ref{PBvsDelta} appears to exist for any
finite
 $\Delta \neq 0$.

This implies that the MI state in $1D$ exists only within the
usual Hubbard model and in ionic Hubbard lattices only in the
limit $U=\infty$. At large $U$ the ionic Hubbard model has been
mapped onto the Heisenberg spin model.  The present finding
suggests that such a mapping may be insufficient for Hubbard
models with ionic potentials and that terms ignored or considered
small possibly play a fundamental role.

\vspace{0.25in}
\vspace{0.25in}

%RMM - renoved paragraph that was repetition

\section{Long Range Order as Inferred by Bond Order Correlation
Function}

\begin{figure}
\hspace{-0.25in} \centering\begin{minipage}[H]{3.5in}
\epsfig{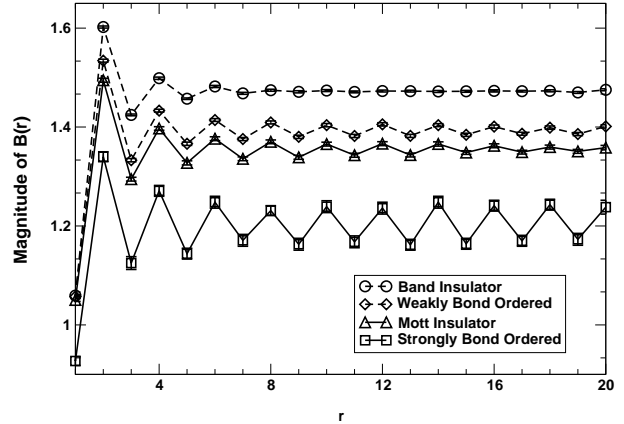}
\caption{Bond order correlation functions for i) band insulating
ii) weakly bond ordered iii) mott insulating and iv) strongly
bond ordered regimes. See text.} \label{BondOrderCorrFuncts}
\end{minipage}
\end{figure}

Existence of a long range bond ordered state can be inferred by
measuring the bond order correlation function ($g_{B}(r)$).  We
define $g_{B}(r)$ as
\begin{equation}
g_{B}(r)=\frac{1}{L}\langle\sum_{i}\hat{B}_{i}\hat{B}_{i+r}
\rangle \label{BOCorrelationFunction}
\end{equation}
where $\hat{B}_{i}$ is defined in Eq.~\ref{Dimerization} and is
the strength of the $i^{th}$ bond of the lattice.  If the BO state
exists then this correlation function would be staggered as a
consequence of the periodic arrangement of dominant and weak
bonds. In Fig.~\ref{BondOrderCorrFuncts} $g_{B}(r)$ is plotted for
$4$ separate cases: (i) $\Delta=0.5714$, $U=1.2$, $L=60$ (ii)
$\Delta=0.1432$, $U=2.4$, $L=122$ (iii) $\Delta=0$, $U=2.5$,
$L=122$ and (iv) $\Delta=0.5714$, $U=3.45$, $L=60$.  The first
case corresponds to the band insulating regime in which $g_{B}(r)$
exponentially approaches a constant, confirming the lack of any
long range ordered phase.  Conversely in the last case, which
corresponds to the system in Fig~\ref{PhaseSep} that exhibited
phase separation, its clearly visible that $g_{B}(r)$ is
staggered, signifying the presence of a long range bond ordered
state. Finite size effects in each of these cases were determined
to be miniscule and small systems were deemed sufficient to
measure $g_{B}(r)$.  The second case corresponds to diminishing
$\Delta$ so as to move the system towards the established Mott
State of the usual Hubbard model. At this point in the phase
diagram the wells of the bimodal distribution are weakly defined;
thus making it extremely difficult to observe the phase separation
of the bond order parameter directly.  In contrast, the bond order
correlation function is clearly staggered, though to a lesser
degree than that of the later case.  Case iii) is the Mott state
of the usual Hubbard model. The staggered behavior of $g_{B}(r)$
does not approach a finite limit at large $r$, but rather tends to
$0$ in a fashion that appears to be a power law ; contrary to the
exponential convergence observed in the BI regime. Comparison of
cases ii) and iii) shows that in each case the staggered behavior
of $g_{B}(r)$ is longer ranged than in the band insulating and
strongly bond ordered cases; this exemplifies the difficulty of
measuring the bond order or the phase separation of this
generalized model as the ionic potential tends to zero.
%TJ
\vspace{0.25in}
\vspace{0.25in}
%TJ
The staggered nature of $g_{B}(r)$ can be used to determine the
mean bond order of the lattice.  Defining two new variables
\begin{equation}
\Delta g_{B}(r) = [ g_{B}(r) - g_{B}(r+1) ] \cdot (-1)^{r}
\end{equation}
and
\begin{equation}
\overline{g_{B}}(r) = \frac{g_{B}(r) + g_{B}(r+1)}{2}
\end{equation}
and substituting Eq~\ref{BOCorrelationFunction} in place of
$g_{B}(r)$ we can relate these two quantities in terms of
measurable quantities.  At large $r$ the $i^{th}$ and $j^{th}$
bonds are uncorrelated and $\Delta g_{B}(r)$ is the RMS bond
order of the lattice whereas $\overline{g_{B}}(r)$ is the square
of the average bond strength.

%TJ2
  In the limit of large $r$ the
average bond order $\langle \hat{B} \rangle$ can be expressed in
terms of $\Delta g_{B}(r)$ as:
\begin{equation}
\sqrt{2\;\Delta g_{B}(r)}=\langle \hat{B} \rangle(r)\;@>>{r\gg
r_{corr}}>\; \langle \hat{B} \rangle.
\end{equation}
%TJ2

\begin{figure}
\hspace{-0.25in} \centering\begin{minipage}[H]{3.5in}
\epsfig{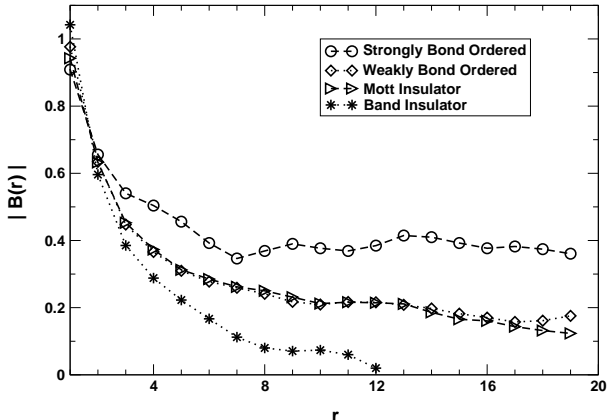} \caption{$\langle
\hat{B} \rangle(r)$ vs $r$ for i) band insulating ii) weakly bond
ordered iii) mott insulating and iv) strongly bond ordered
regimes. } \label{B_r}
\end{minipage}
\end{figure}

This estimate is exact when $r\gg r_{corr}$ where $r_{corr}$ is
the correlation length. Fig~\ref{B_r} shows $\langle \hat{B}
\rangle(r)$ plotted vs r for the same cases as in
 Fig~\ref{BondOrderCorrFuncts}. The strongly bond ordered system
converges to an estimate of the bond order that is remarkably
close to that obtained by extrapolating from distorted lattices
($0.45$).  The weakly bond ordered case appears to converge to a
value near $0.18$ which is in reasonable agreement with the
extrapolated value of $0.2263(46)$.  The band insulating estimate
rapidly approaches $0$ as a function of $r$, whereas the Mott
insulating system appears to approach $0$ in a power law fashion.
%TJ
\vspace{0.2in}
\vspace{0.25in}

\section{Discussion}

One of the primary results of the present work is the quantitative
demonstration of the stability of the bond-ordered phase for
interaction $U$ above a critical value $U_c(\Delta)$ for any
non-zero $\Delta$. Most of our work was carried out through
dimerizing the lattice by $\delta$ and examining the $\delta
\rightarrow 0$ limit.  There were two reasons for this: (i) this
is an aid in the actual calculations which are stabilized by the
applied bias, and (ii) the variation with dimerization $\delta$ is
important in and of itself. Regarding the second point, it is well
known that the ordinary non-ionic Hubbard model is unstable to
dimerization, with a logarithmic Peierls instability at $U=0$ that
becomes a stronger fractional power law instability at large
$U$\cite{Baeriswyl,Hayden,Horsh,Hirsh,Soos,Black}.  Our work shows
that as a function of $U/\Delta$ the ionic Hubbard model undergoes
a phase transition from a stable non-dimerized BI phase to a
correlated phase in which the instability is {\it more severe}
than in the non-ionic Hubbard model.  This is evident in
comparison of Fig.2  with Figs. 5 and 7. In the former for the
non-ionic case, the decrease of the bond order with $\delta$ is
clearly observed and is consistent with previous theoretical
predictions of the power law form. However, in all the
calculations for the ionic model for $U/\Delta$ above the critical
value, the average bond order $<\hat{B}>$ is found to extrapolate
to a non-zero value.  This is observed even for $\delta$ much
smaller than previous studies.  From this evidence alone there are
two possibilities: 1) the BO phase with broken symmetry is stable
at zero dimerization, or 2) there is non-analytic behavior as
$\delta \rightarrow 0$ which is even stronger than that for the
non-ionic Hubbard model.

This result is sufficient to draw conclusions about real 1D
systems in which the sites are allowed to dimerize if this leads
to lower energy.  In either scenario above dimerization would
always occur (except in the BI phase).  In the former case the BO
phase would occur spontaneously and by symmetry there would
always be an accompanying lattice distortion.  In the second
scenario, dimerization would occur and lead to bond order. The
symmetric Mott insulator would never occur and the only transition
would be from a BI phase to a dimerized BO phase.

%TJ2
At this point we can compare with experiment on 1D materials.
Experimental works by Torrance, et al.,\cite{Torrance} observed a
second order transitions between neutral (BI) and ionic (BO)
states in organic charge transfer solids.  The transition occurs
upon applying pressure over a wide range of temperatures and was
attributed to the rise in Madelung energy of the crystal.  No
state synonymous to the Mott state was observed.
%TJ2

%TJ2
In addition, however, calculations on the centrosymmetric lattice
show directly existence of the BO phase in our simulations.  One
observation is the ``phase separation''  or ``flip-flop' between
left and right BO phases as a function of imaginary time in the
QMC simulations.  The other is the staggered behavior of the BO
correlation functions for which the numerical data out to large
distance supports power law behavior in the non-ionic case and
long range order in the ionic case.
%TJ2

Other work on related systems also has identified BO phases.
Recent work by Nakamura \cite{Nakamura} in the extended Hubbard
model has found a rich phase diagram in which there are $2$
transitions from a $BI\rightarrow BO$ and $BO \rightarrow MI$
regime.  The extended Hubbard model differs from the ionic model
studied here in that there is an additional next nearest neighbor
coulomb potential $V$ and no ionic potential $\Delta$.  Nakamura
identifies the first transition as belonging to the gaussian
universality class and the later as a Kosterlitz-Thouless
transition. The $BO$ phase observed by Nakamura exists at all $V$
down to the usual Hubbard model ($V=0$); where both the gaussian
and the KT transitions coexist.  We can imagine the $U_{c}$ at
which the $BI\rightarrow BO$ transition takes place increasing
concurrently as the ionic potential is increased from zero.

Let us now consider why the BO phase was not found in previous
studies that used exact diagonalization Lanczos techniques to
treat small finite systems\cite{RestaSorella,Ortiz}. There are
two reasons why these studies did not find the BO phase.  In the
BO phase the energy can parameterized by the bond order parameter
(Eq~\ref{Dimerization}) that develops a bimodal distribution with
minima at $\pm B$. The true ground state is a linear combination
of these $2$ degenerate BO states,
\begin{equation}
\Psi_{0}=\cos(\theta)\Psi_{+} + \sin(\theta)\Psi_{-},
\label{LanczosGroundState}
\end{equation}
which of course has no net bond order. The situation is similar in
many aspects to a ferromagnet; it is only in the thermodynamic
limit that one or the other of the two states is the true ground
state with long range order.  For finite systems existence of the
BO state can be inferred from correlation functions; however, to
our knowledge this has not been done in other work.

A second reason that the BO states have not been observed may be
that there is no bimodal distribution for the small cells studied
by exact diagonalization. We have addressed this issue using QMC
by measuring the average bond order on distorted lattices of
$14\leq L\leq 62$ sites and extrapolating to the centrosymmetric
limit. At this point in the phase diagram ($\Delta = 0.0716$, $U =
2.4$) lattices with less than $50$ sites do not exhibit the BO
phase and only upon working with larger supercells does QMC detect
the phase separation of the two BO states.  Consequently, exact
diagonalization methods are not currently feasible in such cases
since they scale exponentially with $L$.

Recent work using the density matrix renormalization group (DMRG)
method has reported results for charge $\Delta_c$ and spin
$\Delta_s$ gaps in these models.  This approach should enable one
to distinguish the phases since (i) $\Delta_c =  \Delta_s \neq 0$
in the BI phase, (ii)  $\Delta_c \neq  \Delta_s \neq 0$ in the BO
phase, and (iii) $\Delta_c \neq 0$ but $\Delta_s = 0$ in the MI
phase.  It was found to be very difficult and to require
extremely large cells to determine spin gaps in the BO/MI phases,
and the two reports came to opposite conclusions on the existence
of the BO phase.
%TJ2
  In our QMC calculations we have also determined
the charge and spin gaps.  Our estimates of the charge gap are in
qualitative agreement with other works; however, the spin gap is
very small in all cases except in the BI regime and statistical
noise does not permit accurate determination of such small gaps
in QMC.
%TJ2

Both DMRG calculations find the spin gap to vanish, i.e., the MI
phase to be the ground state for large $U$. We have no direct
explanation of this difference: it may be that our procedure is
not sufficiently accurate to determine the BO-MI transition,
which is the most difficult part of the present work. On the
other hand, it may be that the DMRG calculations on finite cells
with open boundary conditions may have difficulties: the surface
effects break the symmetry of the problem which may lead to
extremely problematic size effects and potential errors.  In any
case, we are very confident that our work establishes that the BO
state is either the ground state or very close to the ground
state in energy; this is clear from our tests on the ordinary
Hubbard model shown in Fig. $1$.

%Does this imply
%that the usual Hubbard model is unstable to bond ordering upon
%making $\Delta\neq 0$?

\section{Conclusions}

We  have studied the phase diagram of an idealized dielectric, the
$1$D ionic Hubbard model proposed by Nagosa\cite{Nagosa} and
Egami\cite{Egami}.  This model undergoes a phase transition as a
function of the on-site interaction $U$, which has been a source
of controversy.  The only previous quantitative
studies\cite{RestaSorella,Ortiz} concluded that at a critical $U$
there is an abrupt "topological" transition from a band insulator
to a Mott insulator with no broken symmetry or long range order
in either phase.  The signature of the transition was found to be
an abrupt change of $1/2$ in the polarization at which the
effective charge diverged signifying the delocalization of the
electron states\cite{RestaSorella,Ortiz}. Recently, however,
there has been a prediction\cite{Fabrizio} that this model would
exhibit two quantum phase transitions: the first signifying a
change of state from a band insulator to a broken symmetry phase
with long range alternating bond order, and the second a
transition to the Mott insulator.

We have studied this model using quantum Monte Carlo methods which
allow the simulation of much larger systems than studied by exact
diagonalization\cite{RestaSorella,Ortiz}. To our knowledge, this
is the first application of QMC to determine the polarization and
localization of an electronic system. We evaluate the expectation
values of the bond-order, polarization and localization using the
expressions Eqs ~\ref{Dimerization}, ~\ref{Polarization} and
~\ref{Localization}.  It is found that upon crossing a critical
value $U_{c}$ a change of phase occurs from a band insulating to
bond-ordered state.  The bond order develops continuously (See
Fig~\ref{CriticalFit}) as a function of $U - U_{c}$ and since the
inversion symmetry is broken,  the polarization also varies
continuously, unlike the results of the small cell
exact-diagonalization calculations.\cite{RestaSorella} The
critical behavior is uniquely determined by fitting the bond
order and polarization to a scaling function near the critical
regime. We find an exponent near 1/2, which differs from that for
the Ising class proposed in Ref~\cite{Fabrizio}; however, it may
be that we are outside of the regime in which the scaling belongs
to the appropriate universality class.  In addition, we found that
there is a metallic point at $U_{c}$ where the system is
metallic.  At this point the charge gap must vanish which we have
found in
%TJ2
 pure ground state calculations by determining the
%TJ2
fluctuations of the polarization.  The calculations determine
quantitatively the localization
length\cite{RestaSorella,Aligia,Souza}, which diverges at the
transition.

An important part of the present QMC work is that we use a forward
projection scheme which allows exact estimates, in principle, of
any operator, including ones such as the polarization (or center
of mass position operator) that do not commute with the
Hamiltonian. Furthermore the nodes of this $1$D  model are known
exactly, so the QMC method is in principle exact. In order to
confirm  the existence of the bond ordered state, we carried out
calculations on dimerized lattices ($t_{o}\pm \delta$) whose
inversion symmetry is explicitly broken, and let $\delta$ become
small. QMC allows us to work with large enough lattices so as to
study systems with levels of dimerization much smaller than
previously feasible\cite{Baeriswyl,Hayden,Horsh,Hirsh,Soos}.  We
find good agreement with previous results obtained from the
Heisenberg spin model that predict electronic correlation
enhances the instability to bond ordering.\cite{Black,Soos}  In
addition, we can see from the simulations of the symmetric
lattice that the system is alternating between the two degenerate
states of bond-order (see Fig~\ref{PhaseSep}).

We have searched for the proposed transition to a Mott insulating
state, but we have not observed such a transition from the bond
ordered regime even for very large $U$ or very small $\Delta$.
Even the smallest value of $\Delta$ considered in this study
($\Delta/t_{o}=1/14 \ll U/t_{o}=2.4$) is sufficient to cause the
ionic Hubbard model to be unstable to bond ordering, although
there is no broken symmetry in the usual Hubbard model ($\Delta
\equiv 0$), neither in the exact solution\cite{Lieb} nor in our
results. Thus our results show that the instability to
dimerization is even stronger in the ionic model than that known
previously for the ordinary non-ionic Hubbard
model.\cite{Baeriswyl,Hayden,Horsh,Hirsh,Soos,Black}. Furthermore,
for the centrosymmetric lattice  ($\delta = 0$), calculations of
correlation functions and observations of ``flip-flop'' between
left and right bond-ordered states in the QMC simulations provides
further evidence for the stability of the bond-ordered state.

Among the interesting consequences of the stability of the BO
state is the existence of fractional
charges.\cite{MeleRice,Fabrizio}  For the case of a dimerized or
bond-ordered state, the charge is an irrational fraction the value
of which depends upon the value of
$\Delta$\cite{MeleRice,Fabrizio}.

Finally, these results imply that if dimerization is allowed
(which is always the case in real materials since the atoms can
always dimerize if it lowers the energy) then the symmetric Mott
state is never stable and the only phase transition is from the
symmetric BI to the dimerized BO state.
%TJ2
This is experimentally confirmed by Torrance et
al.\cite{Torrance}; where upon increasing the electronic
interaction a $BI \rightarrow BO$ transition takes place.

\section{Acknowledgements}

We gratefully acknowledge Erik Koch for his help implementing
lattice quantum Monte Carlo methods and David Campbell, Michele
Fabrizio, Eduardo Fradkin, Erik Koch, Gerardo Ortiz, Rafaelle
Resta, Anders Sandvik, Pinaki Sengupta, Ivo Souza and David
Vanderbilt for invaluable discussions. This work would not have
been feasible without the computational resources of the National
Supercomputing Center at the University of Illinois and the
Materials Research Laboratory. We are especially grateful to the
NT supercluster, managed by Rob Pennington and Michael Showerman,
which provided the majority of the computational resources for
this study. This work was supported by the National Science
Foundation Grant No. DMR $9802373$.

\begin{table}
\begin{tabular}{cccc}
 & A & $U_{c}$ & $\xi$  \\ \tableline
$P$  & 0.44(1) &  2.60(5) & 0.60(10) \\ $\hat{D}$ & 0.49(1) &
2.65(2) & 0.39(4) \\\tableline
\end{tabular}
\label{FirstTable} \caption{Fitting parameters for polarization
and bond order. The quantities in parenthesis are the error in the
last decimal place.}
\end{table}

\end{document}